\documentclass{iopjournal}

\usepackage{lineno}
\usepackage{amsmath,amssymb,amsfonts}
\usepackage{fancyhdr} 
\makeatletter
\renewcommand{\articletype}[1]{{\vspace*{-8mm}\noindent
\Large \sf Machine Learning: Science and Technology}

\vspace*{8mm} \noindent\reversemarginpar
\marginpar{\vspace{-3mm} {\color{gray}\hrule} \ \\ Crossmark\\
{\color{gray}\hrule} \ \\ \tiny {\sf RECEIVED} {\small \\ dd Month yyyy}\\ \\
{\sf REVISED} {\small \\ dd Month yyyy}}
{\scriptsize \sf{\bfseries \MakeUppercase{#1}}}}
\makeatother

\usepackage{cite}

\begin{document}
\articletype{Paper}



\title{\bf Fast reconstruction-based ROI triggering via anomaly detection in the CYGNO optical TPC}

\author{
F D Amaro$^1$,
R Antonietti$^{2,3}$,
E Baracchini$^{4,5}$,
L Benussi$^6$,
C Capoccia$^6$,
M Caponero$^{6,7}$,
L G M de Carvalho$^8$,
G Cavoto$^{9,10}$,
I A Costa$^6$,
A Croce$^6$,
M D'Astolfo$^{4,5}$,
G D'Imperio$^{10}$,
G Dho$^6$,
E Di Marco$^{10}$,
J M F dos Santos$^1$,
D Fiorina$^{4,5}$,
F Iacoangeli$^{10}$,
Z Islam$^{4,5}$,
E Kemp$^{11}$,
H P Lima Jr$^{4,5}$,
G Maccarrone$^6$,
R D P Mano$^1$,
D J G Marques$^{4,5}$,
G Mazzitelli$^6$,
P Meloni$^{2,3}$,
A Messina$^{9,10}$,
V Monno$^{9,10}$,
C M B Monteiro$^1$,
R A Nobrega$^8$,
G M Oppedisano$^{4,5,*}$,
I F Pains$^8$,
E Paoletti$^6$,
F Petrucci$^{2,3}$,
S Piacentini$^{4,5}$,
D Pierluigi$^6$,
D Pinci$^{10}$,
F Renga$^{10}$,
A Russo$^6$,
G Saviano$^{6,12}$,
P A O C Silva$^1$,
N J Spooner$^{13}$,
R Tesauro$^6$,
S Tomassini$^6$,
D Tozzi$^{9,10}$
}

\affil{$^1$ LIBPhys, Department of Physics, University of Coimbra, 3004-516 Coimbra, Portugal}
\affil{$^2$ Dipartimento di Matematica e Fisica, Università Roma Tre, 00146 Roma, Italy}
\affil{$^3$ INFN Sezione di Roma Tre, 00146 Roma, Italy}
\affil{$^4$ Gran Sasso Science Institute, 67100 L'Aquila, Italy}
\affil{$^5$ INFN Laboratori Nazionali del Gran Sasso, 67100 Assergi, Italy}
\affil{$^6$ INFN Laboratori Nazionali di Frascati, 00044 Frascati, Italy}
\affil{$^7$ ENEA Centro Ricerche Frascati, 00044 Frascati, Italy}
\affil{$^8$ Universidade Federal de Juiz de Fora, Faculdade de Engenharia, 36036-900 Juiz de Fora, MG, Brasil}
\affil{$^9$ Dipartimento di Fisica, Sapienza Università di Roma, 00185 Roma, Italy}
\affil{$^{10}$ INFN Sezione di Roma, 00185 Roma, Italy}
\affil{$^{11}$ Universidade Estadual de Campinas (UNICAMP), Campinas 13083-859, SP, Brazil}
\affil{$^{12}$ Dipartimento di Ingegneria Chimica, Materiali e Ambiente, Sapienza Università di Roma, 00185 Roma, Italy}
\affil{$^{13}$ Department of Physics and Astronomy, University of Sheffield, Sheffield S3 7RH, UK}
\affil{$^*$ Author to whom any correspondence should be addressed.}

\email{giuseppe.oppedisano@gssi.it}


\begin{abstract}
Optical-readout Time Projection Chambers (TPCs) produce megapixel-scale images whose fine-grained topological information is essential for rare-event searches, but whose size challenges real-time data selection. We present an unsupervised, reconstruction-based anomaly-detection strategy for fast Region-of-Interest (ROI) extraction that operates directly on minimally processed camera frames. A convolutional autoencoder trained exclusively on pedestal images learns the detector noise morphology without labels, simulation, or fine-grained calibration. Applied to standard data-taking frames, localized reconstruction residuals identify particle-induced structures, from which compact ROIs are extracted via thresholding and spatial clustering. Using real data from the CYGNO optical TPC prototype, we compare two pedestal-trained autoencoder configurations that differ only in their training objective, enabling a controlled study of its impact. The best configuration retains $(93.0 \pm 0.2)\%$ of reconstructed signal intensity while discarding $(97.8 \pm 0.1)\%$ of the image area, with an inference time of $\sim$25 ms per frame on a consumer GPU. The results demonstrate that careful design of the training objective is critical for effective reconstruction-based anomaly detection and that pedestal-trained autoencoders provide a transparent and detector-agnostic baseline for online data reduction in optical TPCs.
\end{abstract}




\section{Introduction}

\begin{figure}[t]
    \centering
    \includegraphics[width=0.32\linewidth]{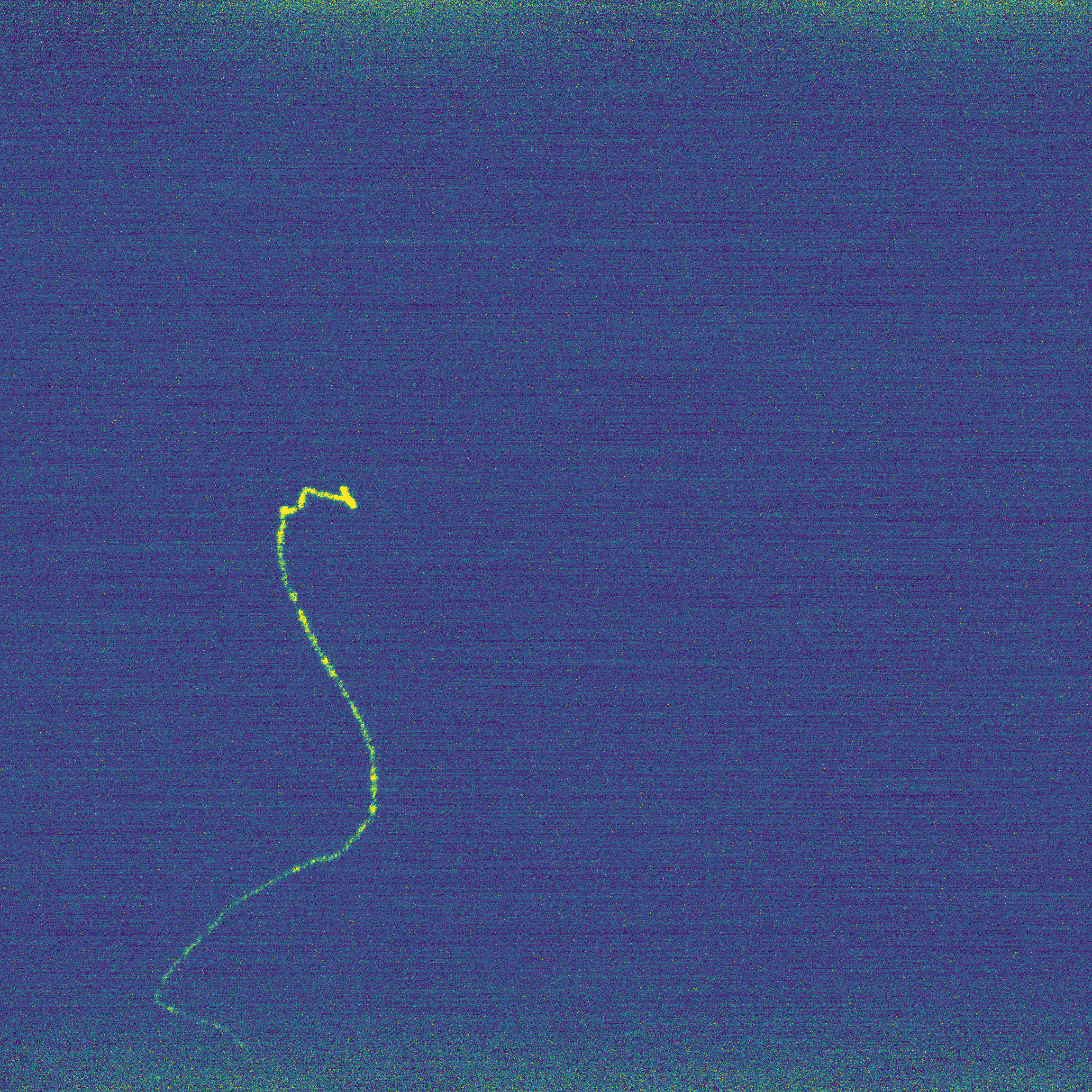}
    \includegraphics[width=0.32\linewidth]{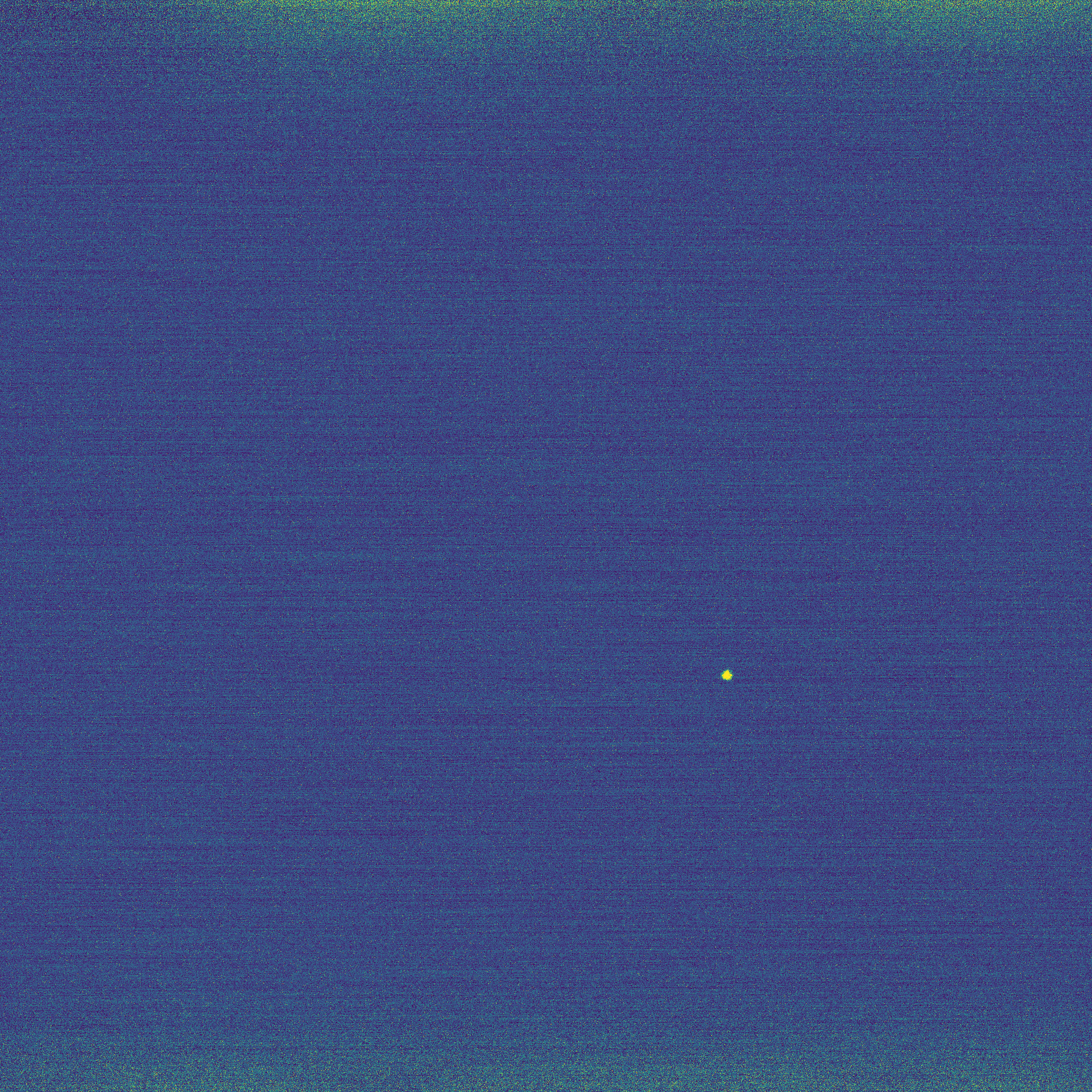}
    \includegraphics[width=0.32\linewidth]{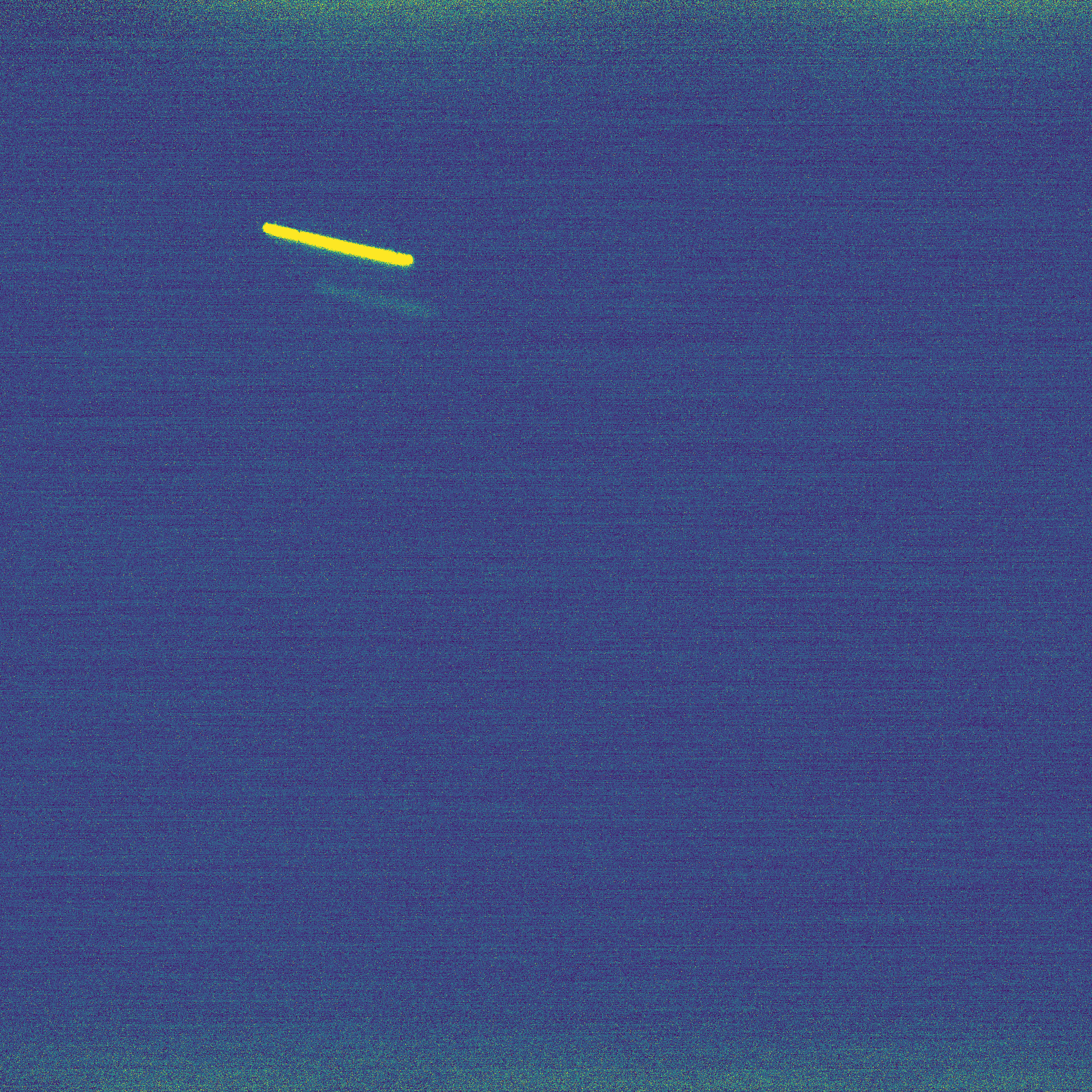}
    \caption{Representative raw sCMOS frames from the CYGNO optical TPC. No ROI selection or processing is applied here, to illustrate the appearance of signals in unprocessed data.}
    \label{fig:raw_events}
\end{figure}

Optical-readout Time Projection Chambers (TPCs) are increasingly relevant tools for rare-event searches in the $\mathcal{O}$(1--100~keV) regime, where short nuclear-recoil tracks---as expected in dark-matter interactions---must be detected amid abundant electronic-recoil backgrounds. In the CYGNO experiment~\cite{Amaro:2022gub}, ionization electrons drift through a He--CF$_4$ gas mixture and undergo charge amplification in a triple-GEM (Gas Electron Multiplier) stack. The resulting CF$_4$ electroluminescence is recorded by scientific CMOS (sCMOS) cameras, yielding finely resolved two-dimensional projections of recoil tracks, complemented by PMT waveforms that enable three-dimensional reconstruction~\cite{Amaro:2025ssv}. This optical readout provides high granularity, low noise, and excellent sensitivity to $\mathcal{O}$(keV) energy deposits, making it attractive for direction-sensitive rare-event searches. 
Representative raw event images are shown in Fig.~\ref{fig:raw_events} to illustrate the typical signal morphology in optical-readout TPC data.

A central challenge of this approach is the size and rate of the raw images. Each exposure covers the entire active volume, producing megapixel-scale frames even in current prototypes. The planned CYGNO-04 demonstrator---a $0.4\,\mathrm{m}^3$ TPC equipped with two $50\times 80\,\mathrm{cm}^2$ optical planes---will acquire $4096\times 2304$-pixel images stored as 16-bit grayscale frames (unsigned 16-bit integer per pixel), corresponding to approximately 18.9\,MB per frame from six sCMOS cameras operating at $\sim$3~Hz, resulting in a data flux of roughly 18 images per second, exceeding 340\,MB/s. Yet the physical signal of interest typically occupies only a few $\mathrm{mm}^2$ in each frame.
Without an online selection mechanism, this would translate into storing enormous volumes of mostly empty data. Any practical data-reduction strategy must therefore sustain at least the full camera rate ($\sim$18 frames s$^{-1}$) so that frames can be filtered on the fly and only physically relevant regions are retained. For sparsely populated megapixel images, retaining only the informative Regions of Interest (ROIs) can dramatically reduce storage and bandwidth requirements. Traditional reconstruction pipelines deliver high-fidelity track characterization~\cite{Amaro_2023} but are too slow for use at trigger level, with typical per-frame processing times of the order of seconds, well above the $\sim$50~ms latency budget required for real-time operation.

Machine learning provides a natural pathway toward fast data selection, especially through unsupervised anomaly detection (AD) techniques that have gained traction across high-energy physics~\cite{varen1, varen_imp, varen3, varen4, varen5,2025arXiv250313195H, 2021JHEP...06..161F,PhysRevD.105.095004, 2022ScPP...12...45O, CMSECAL:2023fvz}. 
In many existing applications, however, reconstruction-based anomaly detection is primarily evaluated at the level of global anomaly scores, while its impact on spatial localization and downstream ROI quality is often not explicitly addressed.
In imaging domains, reconstruction-based AD using autoencoders (AEs) offers a practical mechanism to highlight non-standard structures: by learning to reproduce ``normal'' data, the model identifies particle-induced features as localized reconstruction mismatches. Moreover, the role of the reconstruction objective itself is often less explicitly examined than architectural choices, despite its direct influence on the spatial structure of the resulting residual maps.

Optical TPCs provide an especially favorable setting for this approach. These detectors represent a rare example of a detector system in which large samples of truly background-only data are naturally available, enabling fully unsupervised training without simulation, labels, or signal contamination.
Pedestal frames---images acquired with the GEM amplification switched off---constitute an abundant and clean sample of noise-only data. An autoencoder trained exclusively on these pedestal frames naturally learns the detector’s optical and electronic noise morphology, without relying on simulation, labels, or detailed calibration. In this context, the quality of anomaly localization depends not only on the expressive power of the model, but critically on how the reconstruction objective penalizes localized, structured deviations from the learned noise manifold.
When applied to standard data-taking frames, the network produces residuals that sharply delineate particle-induced structures, from which compact ROIs can be extracted by simple thresholding and spatial aggregation of anomalous pixels.

In this work, we present a complete implementation and evaluation of a pedestal-trained, reconstruction-based anomaly detection strategy for optical-readout TPCs, with particular emphasis on understanding how training-objective design, rather than architectural complexity, shapes anomaly localization and ROI quality.
The study is intentionally exploratory: our aim is to establish a transparent and computationally lightweight baseline for ML-assisted online selection in this detector modality, rather than to optimize performance or architectural complexity. Using real data from the CYGNO prototype, we evaluate the
ROIs predicted by the anomaly-detection pipeline on real data against those derived from the established offline reconstruction algorithm, which serves as a high-fidelity physics reference.
Because the approach is fully unsupervised and relies only on pedestal data, it is broadly
applicable to optical-readout detectors and provides a foundation for future ML-driven data-reduction
pipelines in next-generation experiments.
\section{Optical TPC Data as an ML Domain}
\label{sec:ml_domain}

\subsection{Reconstruction-based anomaly detection in HEP}   

Unsupervised anomaly detection has become a widely explored strategy across collider and astroparticle physics, supporting tasks ranging from rare-event searches to online monitoring and data reduction~\cite{varen1, varen_imp, varen3, varen4, varen5}. In imaging detectors, reconstruction-based approaches are particularly appealing: a model is trained to reproduce background-only data, and deviations in the reconstruction reveal the presence of non-standard or signal-like structures.

Autoencoders~\cite{Goodfellow-et-al-2016, bank2023autoencoders} implement this idea directly. Given an input frame $\mathbf{x}\in\mathbb{R}^{H\times W}$, an encoder 
$E_{\phi}$ maps it to a latent representation $\mathbf{z}=E_{\phi}(\mathbf{x})\in\mathbb{R}^{d}$, and a decoder $D_{\theta}$ reconstructs the image,
\begin{equation}
    \hat{\mathbf{x}} = (D_{\theta} \circ E_{\phi})(\mathbf{x}) .
\end{equation}
The network is trained to minimize a reconstruction loss over a dataset of 
background-only images $\mathcal{D}_{\mathrm{normal}}$,
\begin{equation}
    \min_{\phi,\theta} 
    \; \mathbb{E}_{\mathbf{x}\sim\mathcal{D}_{\mathrm{normal}}}
    \big[\,\mathcal{L}_{\mathrm{rec}}(\mathbf{x},\hat{\mathbf{x}})\,\big],
\end{equation}
where $\mathcal{L}_{\mathrm{rec}}$ enforces similarity between the input $\mathbf{x}$ and its reconstruction $\hat{\mathbf{x}}$. For data that conform to the ``normal'' regime learned during training, the autoencoder reproduces the input well, whereas localized deviations in the residual
\begin{equation}
    \mathbf{r}(\mathbf{x}) = |\mathbf{x}-\hat{\mathbf{x}}|
    \quad\text{or}\quad
    (\mathbf{x}-\hat{\mathbf{x}})^2
\end{equation}
highlight anomalous features. In imaging applications, these spatial residual maps provide a direct mechanism for extracting compact ROIs, making reconstruction-based AD a natural candidate for online selection in detectors that produce large, sparsely populated frames.

\subsection{Pedestal Frames as Normal Data for Unsupervised Learning}

In optical-readout TPCs such as CYGNO, an abundant source of background-only data is provided by \emph{pedestal frames}. Under these conditions, the camera captures only the intrinsic optical and electronic noise of the detector: sCMOS readout noise, fixed-pattern sensor structures, residual dark counts, and static optical features. Since no particle-induced ionization is present, pedestal frames constitute a clean measurement of the detector’s baseline response.

Pedestal runs are taken routinely during CYGNO operation, yielding datasets that accurately reflect the noise morphology of the instrument. This makes them ideally suited for unsupervised training: the autoencoder can learn the characteristic structure of the noise distribution directly from data, without requiring simulation, labels, or detailed calibration. When the model is later applied to standard data-taking frames, particle-induced structures naturally appear as localized reconstruction failures, providing a simple and calibration-light mechanism for anomaly detection and ROI extraction in optical-readout TPCs.


\section{Methods}

\subsection{Data and Preprocessing}

The data used in this study were collected under standard operating conditions of the CYGNO optical-readout TPC operated at the INFN Laboratori Nazionali del Gran Sasso (LNGS). Unless specified otherwise, the chamber was filled with a He--CF$_4$ (60/40) gas mixture, with a drift field $V_{\rm drift}= \text{900}$\,V/cm and the GEM stack operated at $V_{\rm GEM}\simeq 440$\,V per foil during data-taking runs. These operating parameters are reported for completeness, as they influence the noise morphology and the contrast between pedestal and signal-containing frames. Two datasets are employed:

\begin{itemize}
    \item \textbf{Pedestal frames} (used for training), acquired with the GEM amplification 
    voltages disabled, containing only optical and electronic noise.
    \item \textbf{Track-containing frames} (used for evaluation), acquired under standard 
    operating conditions and containing particle-induced ionization tracks.
\end{itemize}

A consistent preprocessing pipeline is applied to both datasets to ensure a uniform input representation for the autoencoder.

\paragraph{Fiducialization.}
Raw frames in this dataset were acquired with the LIME prototype optical readout, producing images of size $2304\times2304$ pixels. Raw camera images exhibit non-uniform response and occasional high-amplitude noise near the edges of the sensor. To suppress these effects, each frame is cropped to a $1525\times1525$-pixel fiducial region defined by the bounding box $(x,y,w,h)=(375,375,1525,1525)$. This removes noisy border regions while retaining the central active area in which physical tracks appear.

\paragraph{Pedestal dataset.}
Pedestal frames, recorded with the GEM voltages switched off, contain only intrinsic detector noise: sCMOS readout noise, fixed-pattern structures, residual dark counts, and static optical features. After fiducialization, each pedestal frame is converted to the $[0,1]$ range via a linear rescaling, after which a pixelwise mean image—computed over the pedestal dataset—is subtracted to remove fixed-pattern offsets. Finally, a global min–max rescaling is applied using scalar extrema computed over the full pedestal sample. All preprocessing operations are pixelwise affine transformations. Denoting the raw value at pixel $i$ by $R_i$, the full transformation can be written as
\[
Y_i = a R_i + b_i ,
\]
where $a$ is a global scalar and $b_i$ is a deterministic per-pixel offset. Consequently,
\[
\mathrm{Cov}(Y_i, Y_j) = a^2 \mathrm{Cov}(R_i, R_j),
\]
so the preprocessing cannot introduce cross-pixel correlations and preserves correlation coefficients.

The pedestal dataset used for training contains 105 frames. Each image provides over $10^6$ pixel samples of the detector noise morphology, so the full pedestal sample corresponds to more than $10^8$ pixel observations. Since pedestal noise is highly homogeneous and stationary, and the training objective is purely reconstructive, the effective statistical sample size is dominated by the number of pixel observations rather than by the number of frames. No explicit hot-pixel masking is applied: persistent camera hot pixels appear with a fixed pattern in the pedestal sample and are removed by the mean-subtraction step. Isolated high-residual pixels arising in signal runs, such as occasional GEM-induced hot pixels, may therefore be identified as anomalous, consistent with the unsupervised nature of the method.

\paragraph{Track dataset.}
For evaluation, we use frames acquired with the GEM stack biased at nominal gain ($\sim$440\,V per foil). These images primarily contain electronic-recoil tracks, i.e. ionization signals produced by electrons, in the $\mathcal{O}(1\text{–}100)$ keV energy range. To ensure compatibility with the pedestal-trained model, the same preprocessing steps are applied: normalization to $[0,1]$, fiducial cropping, pixelwise pedestal mean subtraction, and global min--max rescaling using pedestal statistics. This expresses each track-containing frame relative to the noise morphology learned during training, providing a consistent input space for anomaly detection.

\paragraph{Downscaling.}
After preprocessing, all frames are downscaled from 1525×1525 to 1024×1024 pixels using bilinear interpolation. The images are single-channel 16-bit grayscale frames, so the interpolation is applied uniformly to that channel. No additional explicit antialiasing filter is applied prior to resizing. The linear spatial scaling factor is 1024/1525 $\approx$ 0.67. Track-like structures in the dataset typically extend over tens to several hundreds of pixels at the original resolution. Given this moderate reduction factor and the spatial extent of physically relevant features, the downscaling preserves the topology and contrast required for residual-based anomaly detection, while reducing memory usage and enabling stable training. The goal of the autoencoder is not to reproduce pixel-level track morphology, but to distinguish track-like structures from pedestal-like noise; for this purpose, moderate downscaling does not affect ROI-level sensitivity. Future studies using tiling or multi-scale models will allow full-resolution inference without downscaling.

\subsection{Pixelwise Gaussian Baseline}
As a simple reference method, we consider a pixelwise Gaussian anomaly model constructed from pedestal data. For each pixel $(i,j)$, we estimate the mean $\mu_{ij}$ and standard deviation $\sigma_{ij}$ over pedestal frames. For a test image $x$, we compute the standardized residual
\[
z_{ij} = \frac{x_{ij} - \mu_{ij}}{\sigma_{ij}},
\]
and define anomaly regions via thresholding of $|z_{ij}|$. This provides a simple, fast, and fully unsupervised reference against which the benefits of reconstruction-based models can be assessed.

\subsection{Autoencoder Baseline Architecture}

\begin{figure}[t]
    \centering
    \includegraphics[width=0.8\linewidth]{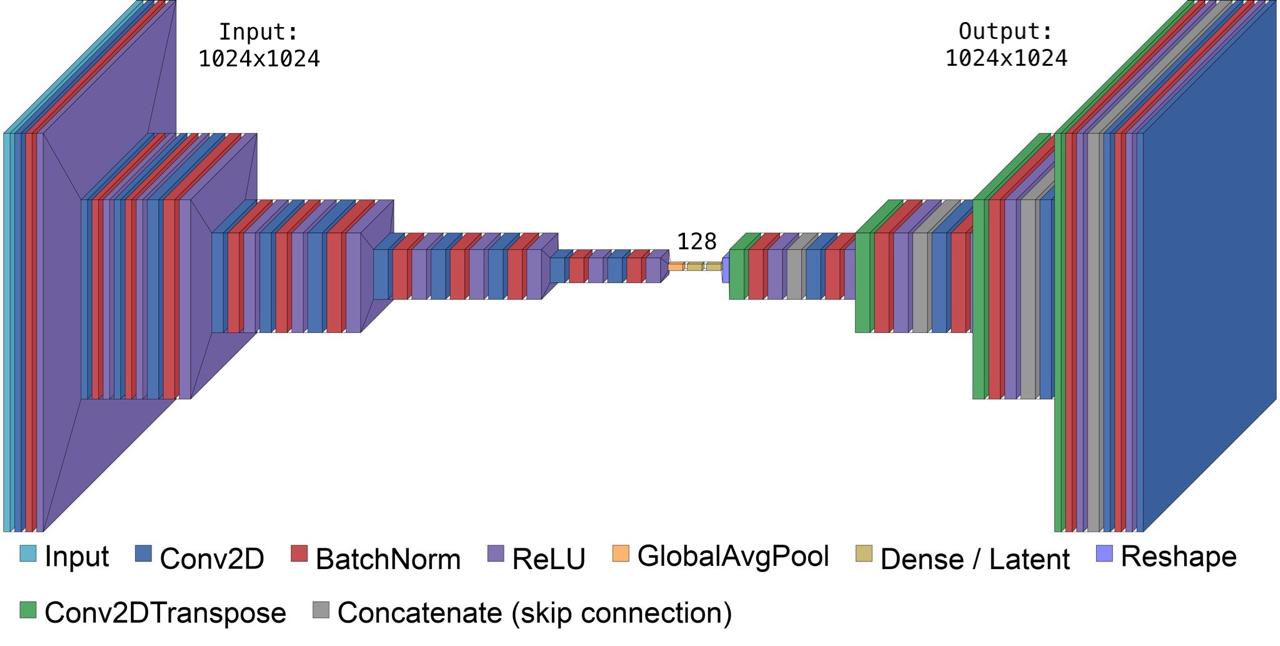}
    \caption{Schematic representation of the convolutional autoencoder architecture. Skip connections are implemented as channel-wise concatenations between encoder feature maps and the corresponding decoder stages at matching spatial resolutions (U-Net--like topology)}
    \label{fig:ae_schematic}
\end{figure}

The anomaly-detection model is a convolutional autoencoder designed to balance reconstruction fidelity with computational efficiency on $1024\times 1024$ images. The network follows a standard encoder--decoder structure (Fig.~\ref{fig:ae_schematic}):

\begin{itemize}
    \item an encoder composed of successive convolutional and down-sampling blocks,
    \item a compact latent representation of dimension 128,
    \item a decoder with transposed convolutions and skip connections, and
    \item a final sigmoid layer producing a normalized single-channel output.
\end{itemize}

\paragraph{Architecture specification.}
To be more precise, the baseline model is a U-Net--like convolutional autoencoder operating on $1024\times1024\times 1$ inputs. The encoder comprises four resolution levels with filter counts $\{22,44,66,88\}$. A \emph{conv block} is defined as a $3\times3$ convolution (stride 1, same padding, no bias), followed by batch normalization and a ReLU activation. A \emph{down block} consists of a $3\times3$ strided convolution (stride 2, same padding, no bias) with batch normalization and ReLU, followed by a conv block with the same number of filters. The spatial resolutions therefore follow $1024 \rightarrow 512 \rightarrow 256 \rightarrow 128 \rightarrow 64$. At the bottleneck ($64\times64\times88$), global average pooling is applied, producing an $88$-dimensional vector, which is mapped to a $128$-dimensional latent representation via a linear dense layer. The decoder is seeded by mapping the latent vector through a dense layer to $64\times64\times3$ and reshaping. Upsampling is performed by four \emph{up blocks}, each consisting of a $3\times3$ transposed convolution (stride 2, same padding, no bias) with batch normalization and ReLU. At each resolution level, the upsampled tensor is concatenated along the channel dimension with the corresponding encoder feature map (skip connection), followed by a $3\times3$ conv block. The final output layer is a $3\times3$ convolution with sigmoid activation producing a single-channel reconstructed image $\hat{\mathbf{x}}$.

Although intentionally simple, this architecture is expressive enough to model the highly homogeneous pedestal noise while remaining computationally efficient for near real-time inference. For general background on these deep-learning components, see 
Ref.~\cite{Goodfellow-et-al-2016}.
All autoencoder variants evaluated in this work share this identical architecture; differences in performance arise solely from changes in the training objective and optimization strategy.

\subsection{Training Objective and Optimization}
\label{sec:loss}

The autoencoder is trained to reproduce the characteristic noise pattern observed in pedestal frames, while failing to accurately reconstruct localized, track-like structures that are absent from the pedestal training dataset and appear only in the evaluation data. In all configurations, training is performed exclusively on pedestal frames, ensuring that the learning process remains fully unsupervised and free from contamination from real signal events.

As a baseline configuration, we employ a hybrid reconstruction loss that combines mean squared
error (MSE) with the Structural Similarity Index (SSIM)~\cite{1284395}:
\begin{equation}
    \label{eq:loss_used}
    \mathcal{L}_{\rm hyb}(\mathbf{x},\hat{\mathbf{x}};\alpha)
    = \alpha\,\bigl(1-\mathrm{SSIM}(\mathbf{x},\hat{\mathbf{x}})\bigr)
    + (1-\alpha)\,\mathrm{MSE}(\mathbf{x},\hat{\mathbf{x}}),
    \qquad \alpha = 0.55.
\end{equation}
The SSIM term is computed using the TensorFlow implementation (\texttt{tf.image.ssim}) with default parameters: an $11\times11$ Gaussian window ($\sigma = 1.5$), constants $K_1 = 0.01$ and $K_2 = 0.03$, and dynamic range $\texttt{max\_val}=1.0$, consistent with the $[0,1]$-normalized input images. The same SSIM configuration is used in both the baseline and refined training objective described below.
SSIM emphasizes local structural agreement, improving the spatial sharpness of the residual maps, while MSE ensures stable global reconstruction of the pedestal noise. The mixing parameter $\alpha=0.55$ controls the relative contribution of the two terms. The value adopted favors the structural similarity term while retaining a non-negligible pixelwise contribution. In practice this choice provides stable convergence and consistent reconstruction of the pedestal noise morphology. This loss provides a simple and transparent reference choice for reconstruction-based anomaly detection.

\begin{figure}
    \centering
    \includegraphics[width=1\linewidth]{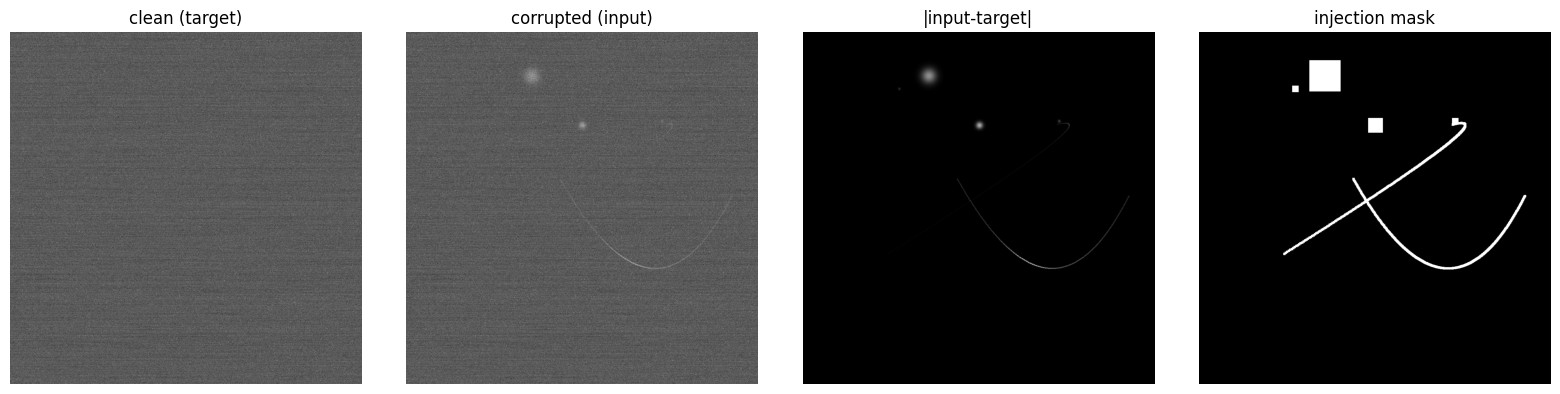}
    \caption{Example of synthetic perturbations injected during training of the refined autoencoder.
From left to right: (a) clean pedestal frame used as reconstruction target; (b) corrupted input obtained by injecting synthetic curved strokes and Gaussian blobs with varying amplitude; (c) absolute difference between input and target (shown for visualization); (d) binary injection mask m used to up-weight the reconstruction loss in perturbed regions. The binary mask marks the conservative support region of each Gaussian deposit as an axis-aligned $\pm 3\sigma$ bounding patch, hence the square patches visible in the mask visualization.
The injected structures are generic and detector-agnostic, and serve solely to regularize the reconstruction objective}
    \label{fig:synthetic-perturbation}
\end{figure}

In addition to this baseline, we explore a refined training configuration designed to reduce the tendency of the autoencoder to partially reconstruct faint, structured deviations. In this variant, synthetic localized perturbations are injected on-the-fly into pedestal frames during training, mimicking generic track- and blob-like structures over a broad range of amplitudes and spatial scales; these perturbations are intentionally uncorrelated with the morphology of real particle tracks (see Fig.~\ref{fig:synthetic-perturbation}). The target reconstruction remains the original, unperturbed pedestal image, so that the network is explicitly trained to suppress the injected structures rather than reproduce them.

To guide this behavior, the reconstruction loss is modified by introducing a spatial weighting term that up-weights the reconstruction error in regions affected by the synthetic perturbations. Concretely, the MSE component of the hybrid loss is weighted by a binary mask identifying the injected regions, while the SSIM term is retained unweighted to preserve global structural fidelity. This procedure does not introduce semantic labels or real signal information: the injected structures are artificial, detector-agnostic, and serve only to regularize the reconstruction objective.

For the refined training configuration, the MSE term is modified by introducing a spatial weight
derived from the synthetic injection mask,
\begin{equation}
\label{eq:weighted_loss}
\mathcal{L}_{\rm ref}
= \alpha \bigl(1-\mathrm{SSIM}(\mathbf{x},\hat{\mathbf{x}})\bigr)
+ (1-\alpha)\,
\bigl\langle (1 + \lambda\,\mathbf{m})\,(\hat{\mathbf{x}}-\mathbf{x})^2 \bigr\rangle ,
\end{equation}
where $\mathbf{m}$ is a binary mask identifying the injected regions and $\langle\cdot\rangle$
denotes the spatial average.

The scale parameter $\lambda$ controls the relative emphasis placed on the injected regions. It was not tuned for performance; instead, it was set using a simple occupancy-based normalization. We estimate the average injected-mask fraction $p=\langle m\rangle$ over a small subset of training batches and choose $\lambda \approx 1/p - 1$, so that, on average, injected regions receive $\mathcal{O}(1/p)$ higher weight than non-injected pixels. In this work we use $\lambda=30$, consistent with the typical mask occupancy observed in the synthetic injections. The loss function is therefore modified to up-weight the reconstruction error in the perturbed regions. This procedure biases the autoencoder toward modeling smooth pedestal fluctuations while deliberately underfitting localized, structured deviations.

\subsubsection{Synthetic perturbation injection and mask generation (refined training)}
In the refined training configuration, each pedestal frame $x$ is converted on-the-fly into a corrupted input $x_{\mathrm{in}} = x + \Delta x$ by injecting a random mixture of localized ``blob'' perturbations and extended ``track-like'' perturbations. The training target remains the original clean pedestal frame $x$. A binary mask $m\in\{0,1\}^{H\times W}$ is generated simultaneously to identify pixels affected by the injection; the mask is used only to up-weight the MSE component of the loss in Eq.~\eqref{eq:weighted_loss}. The injected perturbation $\Delta x$ is additive and independent of the underlying pedestal realization.

\paragraph{Blob perturbations.}
Each blob is a 2D Gaussian added to the image,
\[
\Delta x(u,v)=A\exp\!\left(-\frac{(u-c_x)^2+(v-c_y)^2}{2\sigma^2}\right),
\]
where the center $(c_x,c_y)$ is drawn uniformly over the image, $\sigma\sim U(2,15)$ pixels, and the amplitude $A$ is drawn from a two-component mixture: with probability $p_{\mathrm{faint}}=0.9$, $A\sim U(0.008,0.08)$; otherwise $A\sim U(0.05,0.25)$ (intensities are in the $[0,1]$ normalized scale).

\paragraph{Track-like perturbations.}
A track is constructed by sampling a quadratic B\'ezier curve with three control points $p_0,p_1,p_2$ drawn uniformly over the image. The curve is sampled at $n_{\mathrm{pts}}=300$ locations. At each sample point, a Gaussian deposit is added with width parameter $\sigma=\max(1,w/2.355)$, where $w\sim U(2,6)$ pixels sets the nominal transverse width. The amplitude along the curve is modulated as $A(u)=A_0\,b(u)$ with $u\in[0,1]$, where $A_0$ is drawn from the same faint/bright mixture as for blobs and
\[
b(u)=1+s\exp\!\left[-\frac12\left(\frac{u-u_0}{0.08}\right)^2\right],
\]
with $u_0\sim U(0.2,0.8)$ and $s\sim U(0,2)$.

\paragraph{Number of injected structures and clipping.}
For each training sample, the number of injected structures is drawn as
$n_{\mathrm{tracks}}\sim \mathrm{UnifInt}(1,4)$ and $n_{\mathrm{blobs}}\sim \mathrm{UnifInt}(3,5)$.
After injection, $x_{\mathrm{in}}$ is clipped to the interval $[0,1]$ to match the network input range.

\paragraph{Mask generation.}
For each Gaussian deposit (standalone blob or a deposit along a track) with parameters $(c_x,c_y,\sigma)$, the mask is set to $m=1$ on the axis-aligned pixel patch
$[c_x-3\sigma,c_x+3\sigma]\times[c_y-3\sigma,c_y+3\sigma]$ (clipped to the image boundaries). This produces square mask patches on the pixel grid even though the injected intensity is radially symmetric; the choice is conservative and ensures the weighted-loss region fully covers the injected perturbation.

\subsubsection{Training procedure.}
The refined training configuration modifies only the training objective and data augmentation procedure. The network architecture, input preprocessing, inference pipeline, and evaluation metrics are kept identical to the baseline. Both models were trained using the Adam optimizer~\cite{Kingma2014AdamAM} with initial learning rate $10^{-3}$. A validation split of 10\% of pedestal frames was used for monitoring convergence. Training employed learning rate reduction on validation plateau (multiplicative factor 0.75, patience of 5 epochs) and a batch size of 8.
To ensure a fully converged and fair comparison, both models were trained for 100 epochs with validation monitoring. In practice, validation losses for both configurations reach a clear plateau before the maximum epoch count, indicating stable convergence without evidence of undertraining.

\subsection{Anomaly Scoring and ROI Extraction}
\label{sec:ROI}

After reconstruction, anomaly information is obtained from the pixelwise residual
$\mathbf{r}(\mathbf{x}) = |\mathbf{x}-\hat{\mathbf{x}}|$.
Since the autoencoder is trained exclusively on pedestal frames, the residual map is
characteristically smooth and low-amplitude in noise-dominated regions, while
particle-induced structures appear as localized regions of elevated residuals.
To convert the residual map into compact Regions of Interest (ROIs), we apply the
following procedure:

\begin{enumerate}

    \item \textbf{Residual thresholding.}  
    A global threshold $\tau$ is applied to the residual map to isolate anomalous pixels from the pedestal noise baseline. The threshold is fixed by requiring a strong suppression of residual activity on an independent set of pedestal-only frames. The working point used in this study is $\tau = 0.04$.

    On a held-out sample of 101 pedestal images, this choice results in an average fraction of $6.9\times10^{-3}$ of pedestal pixels being retained within the ROI masks after the full preprocessing and extraction pipeline. When excluding a 50-pixel-wide border region on each side of the image, the retained fraction decreases to $2.2\times10^{-4}$, indicating that the residual activity surviving the threshold is largely concentrated near the image boundaries.

    \item \textbf{Spatial aggregation of anomalous pixels.} 
    The thresholded residual map is converted into a binary anomaly mask and subjected to a spatial aggregation step that links nearby anomalous fragments. This is implemented through a morphological closing operation using a circular structuring element with radius $d_{\mathrm{link}} = 40$ pixels. The disk-like element reflects the approximately isotropic transverse light spread of localized ionization clusters in the optical TPC. The chosen linking radius is large compared with the typical fragmentation scale introduced by residual thresholding, yet small compared with the full extent of extended track-like structures, which span hundreds of pixels in a $1024 \times 1024$ fiducial image. This choice merges fragments belonging to the same physical structure without artificially connecting well-separated tracks, thereby reducing fragmentation of elongated or low-contrast features while preserving overall track morphology.

    \item \textbf{ROI mask construction.}  
    After spatial aggregation, the resulting binary mask directly defines the ROIs. In
    this study no explicit minimum-area requirement is imposed, and isolated anomalous
    pixels—if present—are retained as part of the ROI mask, consistent with the
    unsupervised and signal-agnostic nature of the approach.
\end{enumerate}

The final output of the pipeline is therefore a ROI mask that can be applied directly to the original image. Across the evaluation dataset, these masks typically retain only a small fraction of the total image area while enclosing the vast majority of physically relevant signal pixels.

\subsection{Evaluation Protocol}
\label{sec:evaluation_protocol}

The performance of the anomaly-detection pipeline is assessed by comparing the ROIs predicted from the residual maps with the event topology obtained from the
established CYGNO offline reconstruction algorithm~\cite{Amaro_2023}. Although this reconstruction
is not intended for real-time use, it provides high-fidelity identification of signal pixels and
serves as a reliable reference for evaluating the proposed ML-based method. Moreover, this
reconstruction defines the pixels associated with each physical event in standard CYGNO
analyses; ensuring that the ML pipeline retains all pixels identified by the reference
reconstruction is therefore essential to avoid any downstream loss of physics-relevant
information.

A key element of the evaluation is the choice to operate on a \emph{per-event} basis rather than
per-image. A single camera frame may contain multiple independent particle interactions, and
treating the entire image as a single unit would obscure partial successes or failures (e.g.\
capturing two events while missing a third). Evaluating each reconstructed interaction
independently yields a fine-grained and unbiased estimate of the model’s sensitivity.

For the purpose of performance evaluation only, we apply an additional containment requirement
to suppress known reconstruction artifacts near the sensor boundaries. Specifically, each
reconstructed event is required to contain at least one signal pixel located more than 50 pixels
away from the image border. This selection is analogous to standard fiducial-quality cuts used
in offline analyses, and is applied identically to all methods under comparison. It does not
affect training or inference, but ensures that the reference reconstruction used for evaluation
corresponds to well-contained physical topologies.

Three metrics are used to quantify performance:

\begin{itemize}

    \item \textbf{Signal-intensity coverage.}  
    For each reconstructed event, we compute the fraction of pedestal-subtracted signal
    intensity that lies within the predicted ROIs. Since the summed pixel intensity is
    approximately proportional to the deposited energy in the optical TPC, this metric
    provides an energy-weighted measure of how effectively the extraction pipeline retains
    the physically relevant parts of each interaction.

    \item \textbf{Area cut (area reduction).}  
    We define the area cut as the fraction of the image area discarded by the ROI selection,
    \begin{equation}
    f_{\rm cut} = 1 - \frac{A_{\rm ROI}}{A_{\rm img}} ,
    \end{equation}
    where $A_{\rm ROI}$ is the total area covered by the predicted ROI mask(s) and $A_{\rm img}$ is the full image area. Larger values correspond to stronger data reduction (i.e.\ fewer pixels retained), and thus to greater potential savings in data transfer and storage.

    \item \textbf{Inference time.}  
    The average wall-clock time required for the autoencoder forward pass on a single
    $1024\times 1024$ image. This provides a practical estimate of the feasibility of
    deploying the model in real-time or near–real-time data-acquisition environments.

\end{itemize}

Together, these metrics capture both the scientific relevance (signal preservation) and the
computational advantages (area reduction and latency) of the proposed approach, providing a
comprehensive evaluation of its suitability for online selection in optical-readout TPCs.

Within this evaluation framework, we compare three anomaly-scoring approaches that share an identical preprocessing chain, ROI-extraction procedure, and evaluation metrics, differing only in the definition of the residual or anomaly map. All three methods are evaluated under identical conditions, enabling a direct and fair comparison of their signal-retention and data-reduction performance.


\section{Results}

\subsection{Qualitative Behaviour}

\begin{figure}[t]
\centering
\includegraphics[width=1.0\textwidth]{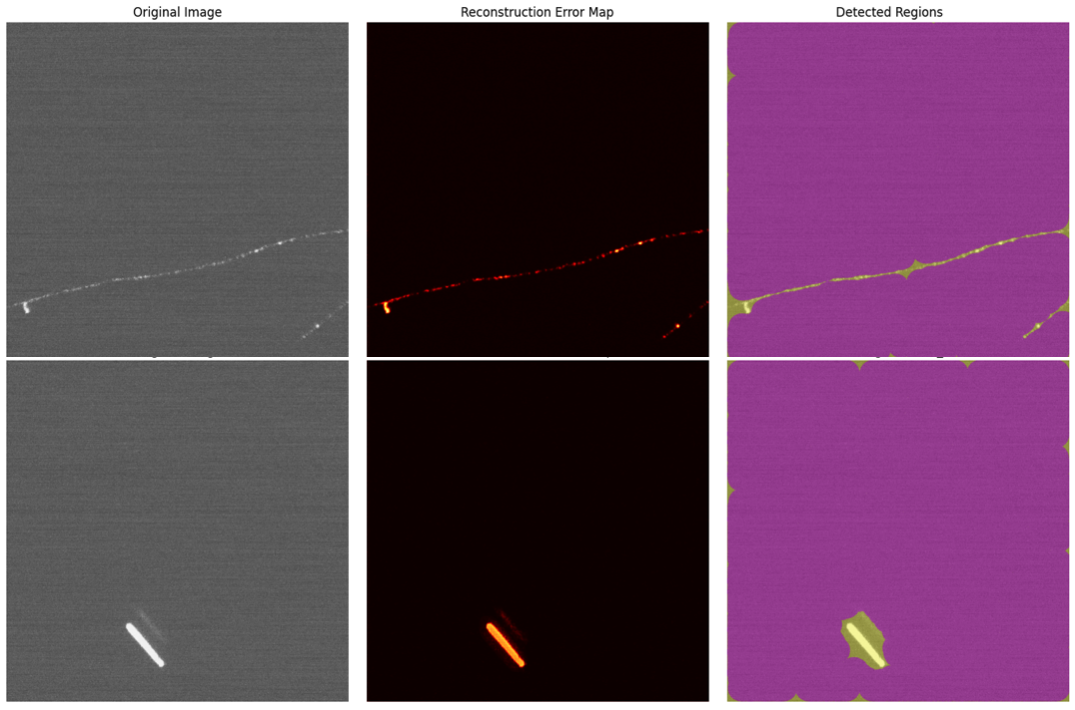}
\caption{Representative Regions of Interest (ROIs) returned by the anomaly-detection framework.
Each row shows: (a) the fiducialized camera image; (b) the anomaly map, where track-like
structures appear as localized high-residual regions; (c) the final ROI mask after spatial aggregation. The ROIs reliably enclose particle-induced
structures while excluding noise-dominated background regions.}
\label{fig:roi_examples}
\end{figure}

The reconstruction-based anomaly-detection framework produces compact ROIs that closely follow the morphology of particle-induced structures. Representative
examples are shown in Fig.~\ref{fig:roi_examples}. For each event, the anomaly map highlights
track-like features with sharp spatial localization, and the subsequent ROI-extraction steps
successfully isolate these regions while excluding the vast majority of noise-dominated
background pixels. Across the full evaluation sample, the predicted ROIs consistently align
with the visible signal structures, illustrating the suitability of residual-based anomaly maps
as a basis for fast and robust localization in optical TPC images.

\subsection{Comparison of Anomaly-Scoring Methods}

We next compare the quantitative performance of the three anomaly-scoring approaches introduced in Section~3: a pixelwise Gaussian baseline, a pedestal-trained autoencoder with baseline training, and a pedestal-trained autoencoder with refined training. All methods are
evaluated on the same event sample, using an identical preprocessing chain, ROI-extraction procedure, and evaluation protocol. Differences in performance therefore reflect only the definition of the anomaly map.

\begin{figure}[t]
\centering
\includegraphics[width=0.95\textwidth]{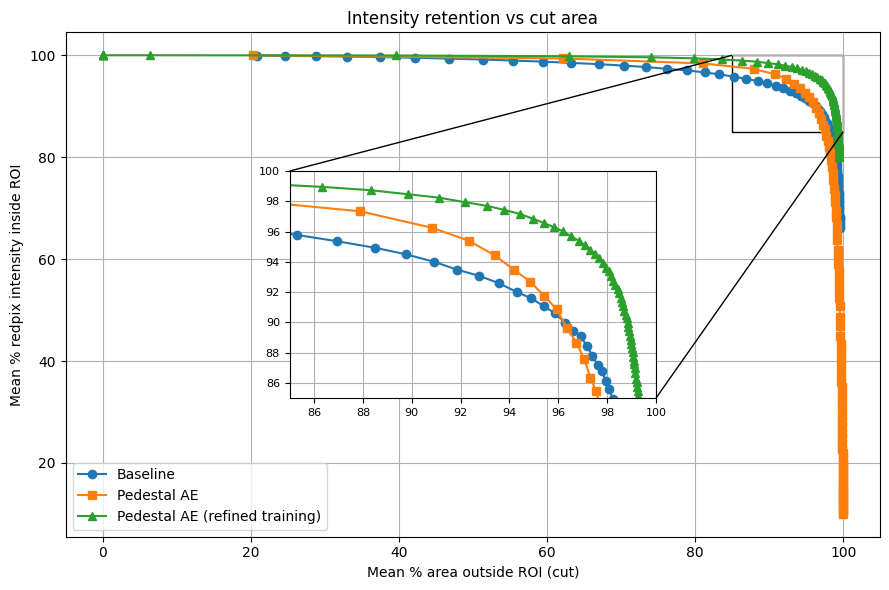}
\caption{Trade-off between mean signal-intensity coverage and mean area cut for the three
anomaly-scoring approaches, obtained by sweeping the residual threshold $\tau$. All methods
share the same ROI-extraction pipeline and are evaluated on the same event sample. Curves
closer to the top-right indicate stronger compression at fixed signal retention.}
\label{fig:tradeoff}
\end{figure}

Figure~\ref{fig:tradeoff} shows the mean signal-intensity coverage as a function of the mean area cut for the three methods, obtained by sweeping the residual threshold $\tau$. Each point corresponds to a single threshold value, and curves closer to the top-right corner indicate a more favorable trade-off between signal retention and data reduction.

The pixelwise Gaussian baseline provides a simple and fast reference and yields a strong trade-off between signal-intensity coverage and area cut across a broad threshold range. The pedestal-trained autoencoder with the baseline reconstruction objective achieves comparable performance and, in the low-compression regime, can provide marginally higher signal-intensity coverage.
The observation that a naïvely trained autoencoder does not outperform a pixelwise Gaussian baseline reflects the tendency of sufficiently expressive autoencoders to partially reconstruct structured deviations that are absent from the training distribution, thereby reducing residual contrast and weakening anomaly separability. Similar behavior has been observed in the anomaly-detection literature, for example, Ref.~\cite{ANGIULLI2026133002, 2021arXiv211009742A}. The refined-training configuration introduced here explicitly mitigates this effect by discouraging reconstruction of localized structured perturbations while preserving pedestal fidelity.
The refined-training autoencoder consistently outperforms both alternatives across the explored threshold range, achieving higher signal retention for a given area cut. Figure~\ref{fig:tradeoff} provides a global view of the signal-retention versus compression trade-off, which offers a more informative representation than single operating-point comparisons and avoids threshold-dependent bias. On the basis of this comparison, the refined-training autoencoder is selected as the reference configuration for the remainder of the analysis.

\subsection{Performance of the Selected Configuration}

We now report detailed performance metrics for the refined-training autoencoder at a fixed operating point, corresponding to the threshold value of $\tau = 0.04$ for the refined autoencoder configuration, as defined in Sec.~\ref{sec:ROI}. The evaluation is performed on 1563 reconstructed events passing the selection described in Section~\ref{sec:evaluation_protocol}.

The selected configuration achieves:
\begin{itemize}
    \item \textbf{Mean signal-intensity coverage:} $(93.0 \pm 0.2)\%$.
    This indicates that the ROIs retain the large majority of the pedestal-subtracted signal
    intensity assigned to each event by the reference reconstruction.

    \item \textbf{Mean area cut:} $(97.8 \pm 0.1)\%$.
    On average, only $(2.2 \pm 0.1)\%$ of the image area is retained, corresponding to a
    reduction of nearly two orders of magnitude in the number of stored pixels.

    \item \textbf{Inference time:} Inference latency was measured from TensorFlow/Keras forward-pass timings with batch size 1 on an Apple M1 Pro (16\,GB unified memory). The reported $\sim$25\,ms per frame corresponds to model inference only, with the network resident in memory; disk I/O and preprocessing are excluded. Due to the unified-memory architecture, no explicit CPU--GPU transfer overhead is incurred in this measurement. No batching or runtime-specific optimizations were applied, so this value represents single-frame inference under standard settings rather than an optimized deployment benchmark. Even under these conservative conditions, the measured latency remains well below the $\sim$50 ms per-frame budget required for online triggering in the CYGNO data-acquisition system, indicating that the approach is compatible with real-time deployment.

\end{itemize}

\subsection{Dependence on Event Energy}
\begin{figure}[t]
\centering
\includegraphics[width=0.9\textwidth]{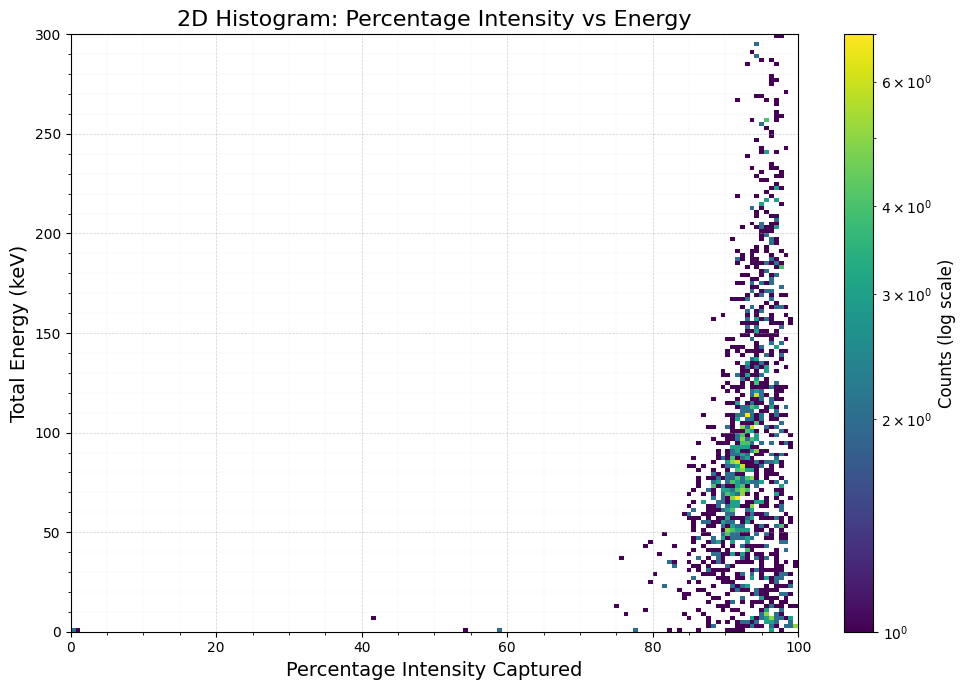}
\caption{Signal-intensity coverage as a function of reconstructed event energy for the
refined-training autoencoder. The method maintains high coverage across the full energy
range. A small number of outliers at very low coverage are discussed separately in
Section~\ref{sec:lowcases}.}
\label{fig:performance_energy}
\end{figure}

Figure \ref{fig:performance_energy} shows the signal-intensity coverage as a function of the reconstructed event energy for the selected configuration. The method maintains high coverage across the full energy range explored, extending from O(keV) events up to several hundred keV.
A mild energy dependence is visible in the distribution: at lower energies the coverage exhibits a somewhat broader spread, while at higher energies the distribution becomes more tightly clustered. The median signal-intensity coverage nevertheless remains high throughout the range, typically between about 90\% and 97\%.
For the intended application as a trigger-level ROI selector, this level of containment is already sufficient for the downstream reconstruction algorithms to reliably process the event. 

\subsection{Visual Inspection of Low-Performance Cases}
\label{sec:lowcases}

\begin{figure}[h]
\centering
\includegraphics[width=0.8\textwidth]{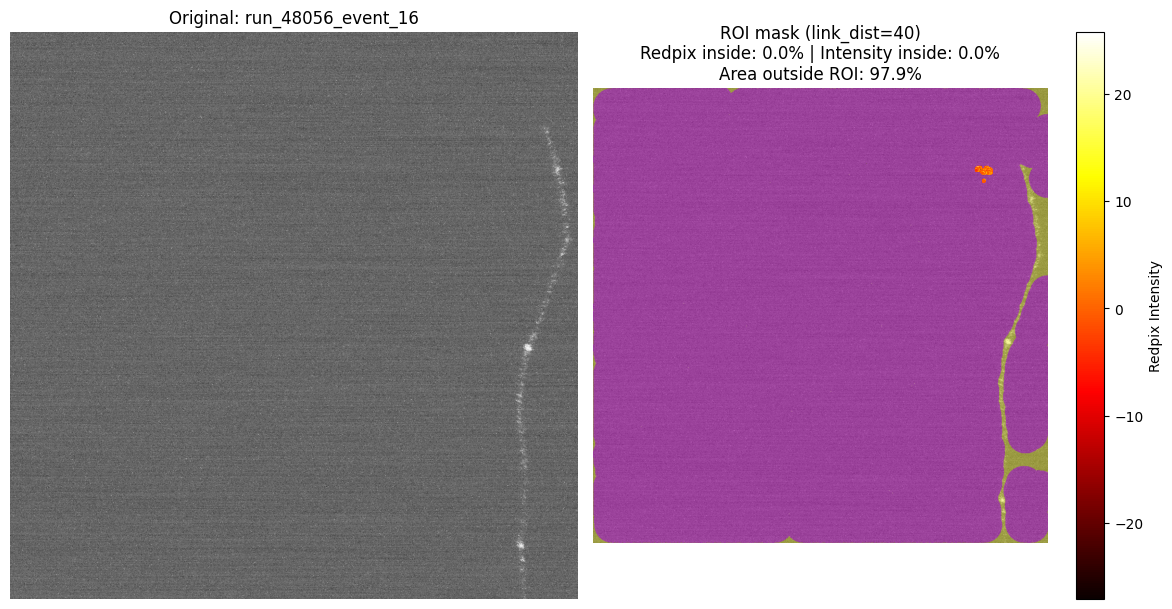}\\[1ex]
\includegraphics[width=0.8\textwidth]{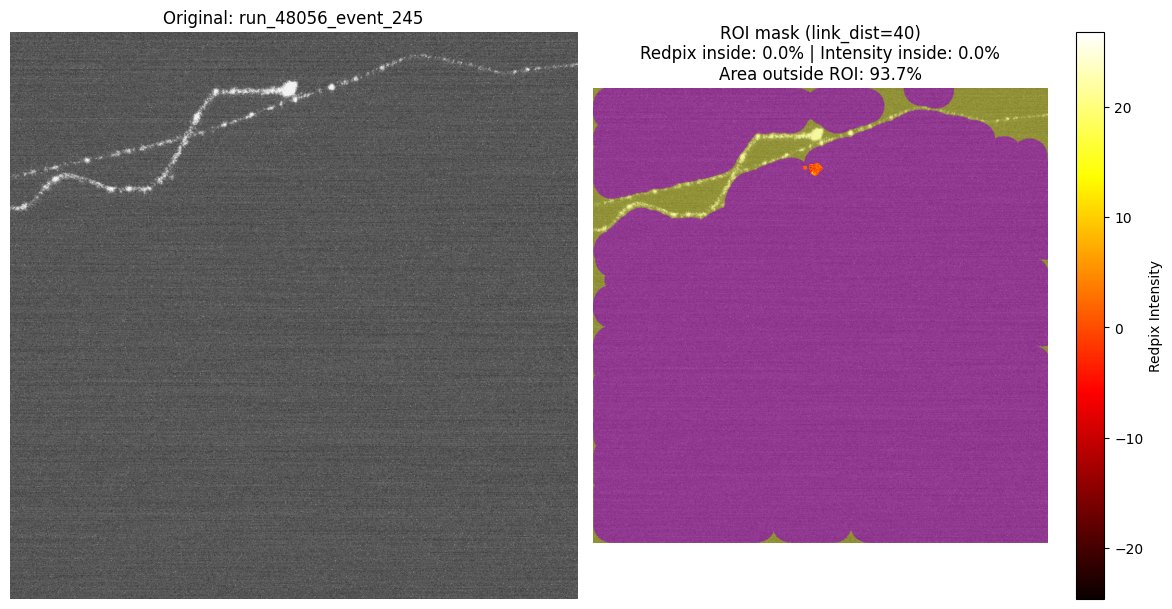}\\[1ex]
\includegraphics[width=0.8\textwidth]{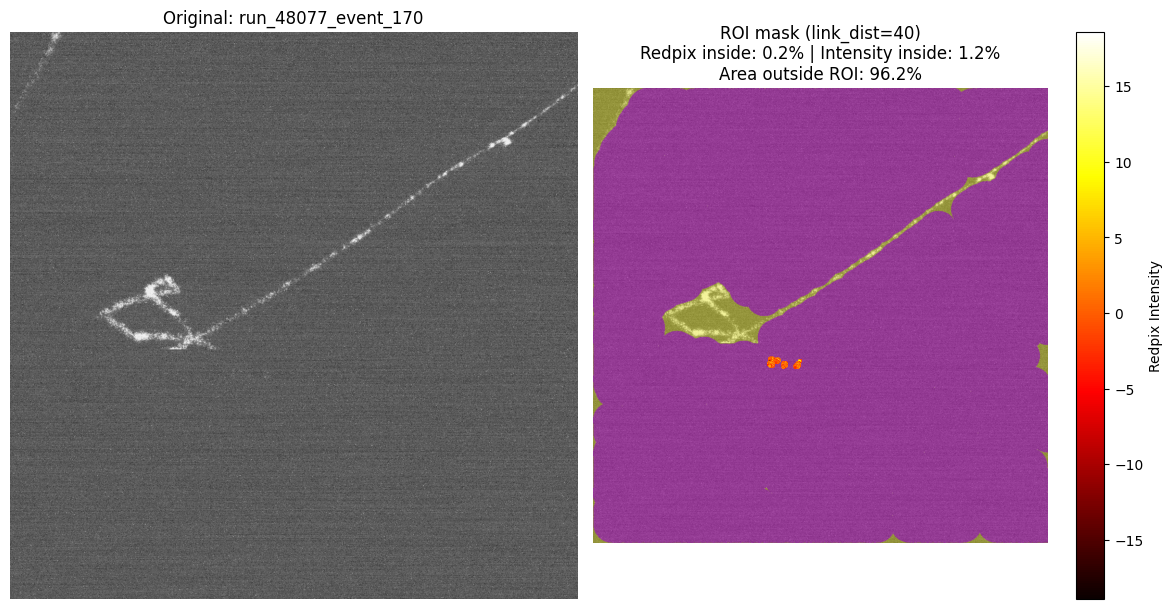}
\caption{Representative examples of events with near-zero signal-intensity coverage. Each panel shows the original camera image and the corresponding ROI mask produced by the refined-training autoencoder, with pixels assigned to the event by the reference reconstruction overlaid. No
clear track-like structures are visible in the raw images, indicating that these cases arise from reconstruction artifacts rather than failures of the anomaly-detection pipeline.}
\label{fig:failure_cases}
\end{figure}

As shown in Fig.~\ref{fig:performance_energy}, the refined-training autoencoder achieves high signal-intensity coverage for the overwhelming majority of reconstructed events. Only a very small number of cases (3 out of 1563) exhibit near-zero intensity coverage. A detailed visual inspection of these events indicates that they do not correspond to genuine failures of the anomaly-detection pipeline.
Representative examples are shown in Fig.~\ref{fig:failure_cases}, where the original camera images are compared directly to the predicted ROI masks overlaid with the pixels assigned to the event by the reference reconstruction to the near-zero signal-intensity coverage events. In all cases, no clear track-like or localized ionization structure is visible in the underlying image. The pixels labeled as signal by the reference reconstruction instead appear as isolated or weak fluctuations consistent with noise, and are not supported by any coherent topology in the camera data.
These events are therefore attributed to artifacts of the offline reconstruction, which may assign spurious pixels or small clusters to an interaction in the absence of a physically meaningful signal. Such low-significance reconstructions are typically removed by quality and fiducial cuts in standard CYGNO analyses. Consistently with this interpretation, all three events have reconstructed energies below $\sim$1\,keV.

From the perspective of an online trigger or data-reduction system, this behavior is entirely acceptable. The anomaly-detection pipeline is designed to retain genuine track-like structures with very high efficiency, rather than to reproduce every pixel assignment of the offline reconstruction. The absence of ROIs in events lacking visible signal structures reflects a conservative and physically sensible response of the model, rather than a limitation of the approach.

\section{Discussion and Perspectives}

The results of this study show that reconstruction-based anomaly detection can provide an effective strategy for fast Region-of-Interest (ROI) identification in optical-readout TPC data, provided that the training objective is carefully defined. Trained exclusively on pedestal frames, the model learns the detector’s characteristic noise morphology and produces residual maps in which particle-induced structures are sharply localized. The resulting ROIs retain nearly all signal-relevant pixels while discarding the large majority of the empty background, offering a practical and calibration-light approach to data-volume reduction prior to downstream reconstruction.

A central outcome of the comparative study is that a simple pixelwise Gaussian model trained on pedestal data provides a strong and non-trivial baseline for anomaly detection in this domain. A pedestal-trained autoencoder with a naïve reconstruction objective does not automatically outperform this classical approach. The refined-training configuration introduced here, which explicitly discourages reconstruction of faint structured deviations, is what enables the autoencoder to surpass the Gaussian baseline in terms of the signal-retention versus area-cut trade-off.

The injected perturbations used in the refined training configuration should not be interpreted as a conventional data-augmentation strategy aimed at enlarging the training dataset. Instead, they act as controlled localized structures used to shape the reconstruction objective and discourage the autoencoder from reproducing track-like patterns. If such perturbations were introduced while retaining a purely reconstructive loss, the network would instead learn to reproduce them, improving reconstruction fidelity but reducing the contrast of localized anomalies in the residual map.

Beyond its quantitative performance, the method has several attractive properties for online selection. The residual maps provide a transparent visualization of discrepancies between each frame and the learned noise model, enabling intuitive diagnostic checks and making failure modes easy to interpret. The computational footprint is modest: $1024\times1024$ images can be processed in tens of milliseconds on a consumer-grade GPU, suggesting that real-time or near–real-time deployment is feasible in future data-acquisition pipelines.

The need for such approaches is amplified by the expected data volumes of the forthcoming CYGNO-04 detector, where multiple high-resolution cameras operating simultaneously will generate raw data streams of several hundred megabytes per second. In this context, isolating only compact regions carrying physically meaningful information is essential for maintaining a sustainable throughput. The present work forms part of a broader machine-learning effort within CYGNO, complementing exploratory studies on event classification and trigger optimization reported in~\cite{oppedisano_thesis}. Together, these developments point toward an increasingly 
ML-assisted online selection framework for future detector stages.

Several practical considerations merit attention. The autoencoder was trained and evaluated on moderately downscaled images; processing full-resolution frames may require tiling strategies or architectural adjustments to remain within memory constraints. As with any reconstruction-based method, extremely faint or diffuse structures may be partially smoothed by the decoder. However, our inspection of low-efficiency cases indicates that such instances are rare and typically arise from imperfections in the reference reconstruction rather than from systematic limitations of the model.

More broadly, this study should be viewed as an initial baseline rather than a final, optimized solution. Optical-readout TPCs present a distinctive data modality—sparse, noise-dominated, and megapixel-scale—that raises interesting challenges for machine learning. Future extensions could include multi-scale autoencoders capable of jointly modeling fine and coarse features, hybrid architectures incorporating attention mechanisms, or uncertainty-aware formulations of the residual maps. Dedicated deployment studies using full-size frames and realistic DAQ conditions will also be crucial to quantify end-to-end throughput and characterize the impact on the global reconstruction chain.

Overall, the results indicate that pedestal-trained autoencoders provide a robust, interpretable, and computationally efficient foundation for ML-driven data reduction in optical-readout TPCs. Although this work focuses on the current CYGNO prototype, the methodology is generic and depends only on the availability of pedestal runs, making it readily transferable to future large-scale CYGNO detectors and to other experiments employing optical gaseous TPC readout.


\section{Conclusion}

We have presented an unsupervised, reconstruction-based anomaly detection framework for fast Region-of-Interest extraction in optical-readout TPC data. Using pedestal frames as a noise-only training sample, we investigated multiple anomaly-scoring strategies within a common preprocessing, ROI-extraction, and evaluation pipeline. We find that while a simple pixelwise Gaussian model provides a strong and competitive baseline, a convolutional autoencoder can surpass this classical approach when its training objective is carefully designed to suppress the reconstruction of faint structured deviations.

Applied to real data from the CYGNO prototype, the refined-training autoencoder achieves high signal-intensity retention while discarding the large majority of empty background pixels, corresponding to an area cut of nearly two orders of magnitude at inference times compatible
with near–real-time operation. These results demonstrate that reconstruction-based anomaly detection offers a practical and transparent solution for
online data reduction in optical-readout TPCs.

The proposed approach is not intended to replace detailed offline reconstruction, but rather to act as an efficient and robust preliminary stage that reduces data volume before more computationally intensive processing. As optical-readout detectors continue to scale in size
and acquisition rate, such ML-assisted strategies are likely to play an increasingly important role in enabling sustainable data acquisition and real-time event selection in next-generation TPC experiments.


\section*{Acknowledgements}

This project has received fundings under the European Union’s Horizon 2020 research and innovation program from the European Research Council (ERC) grant agreement No. 818744 and is supported by the Italian Ministry of Education, University and Research through the project PRIN: Progetti di Ricerca di Rilevante Interesse Nazionale “Zero Radioactivity in Future experiment” (Prot. 2017T54J9J). A. Messina has also been supported by the PNRR MUR project PE0000013–FAIR. We want to thank General Services and Mechanical Workshops of Laboratori Nazionali di Frascati (LNF). We want to thank the INFN Laboratori Nazionali del Gran Sasso for hosting and supporting the CYGNO project.


\bibliographystyle{unsrt}
\bibliography{bibliography}

\end{document}